\documentclass{amsart}
\usepackage[utf8]{inputenc}
\usepackage{comment}
\usepackage{float}
\usepackage{hyperref}
\usepackage{graphicx, overpic}
\usepackage{todonotes}
\usepackage[asymmetric,vmargin=3cm,hmargin=3cm,marginpar=2cm]{geometry}
\usepackage{amsmath,amssymb,amsfonts,amsthm,amscd,graphicx,psfrag,epsfig,fge}
\usepackage[]{hyperref}
\usepackage{mathtools}
\usepackage{thmtools}
\usepackage{tikz-cd}
\usepackage{graphicx,wrapfig,lipsum}

\newtheorem{theorem}{Theorem}

\usepackage{graphicx} 

\title{Joseph Wolstenholme and the Trigonometry of Tetrahedra}
\author{Daniil Rudenko}

\begin{document}

\maketitle






\begin{abstract}
``Mr Wolstenholme was a very old gentleman who came every summer to stay with us. He was brown; he had a beard
and very small eyes in fat cheeks; and he fitted into a brown
wicker beehive chair as if it had been his nest. He used to
sit in this beehive chair smoking and reading. He had only
one characteristic – that when he ate plum tart he spurted
the juice through his nose so that it made a purple stain on
his grey moustache. This seemed enough to cause us perpetual delight. We called him “The Woolly One”. By way
of shading him a little I remember that we had to be kind
to him because he was not happy at home; that he was very
poor, yet once gave Thoby half a crown; that he had a son
who was drowned in Australia; and I know too that he was
a great mathematician. He never said a word all the time
I knew him. But he still seems to me a complete character;
and whenever I think of him I begin to laugh.'' -- Virginia Woolf, {\it Moments of Being.}
\end{abstract}

\

The trigonometry of tetrahedra remains a relatively unexplored and complex area compared to the well-established and extensively studied trigonometry of triangles. While working on Hilbert's third problem, I discovered a new fact in the trigonometry of a tetrahedron: the identity 
\begin{align} \label{FormulaCrossRatios}
\bigl[e^{i\Omega_1},e^{i\Omega_2},e^{i\Omega_3},e^{i\Omega_4}\bigr]=[P_1,P_2,P_3,P_4],
\end{align}
where $[a,b,c,d]=\frac{(a-b)(c-d)}{(a-c)(d-b)}$ is the cross-ratio, $P_i$ are the perimeters of faces, and $\Omega_i$ are the spherical angles of a Euclidean tetrahedron (\cite[Corollary 1.2]{Rud22}). This statement has such a “classical” flavor that I kept wondering if it had been discovered long before my work, see \url{https://mathoverflow.net/q/336464} for the discussion. 
The literature related to the trigonometry of tetrahedra is extensive, yet I was unable to find anything resembling my result anywhere in the published texts. I was at the point of giving up when I came across the name {\it Joseph Wolstenholme} in \cite{Ric02}. Wolstenholme worked on the trigonometry of tetrahedra in the late 19\textsuperscript{th} century. The mathematical community failed to react to his results, likely because they were never published in a peer-reviewed journal and appeared only as problems in the British periodical {\it The Educational Times}. I am going to resist the temptation to state Wolstenholme's theorem at the onset, and will begin instead with some historical context. In doing so, I hope to put this mesmerizing story in perspective.

\section{Joseph Wolstenholme and the trigonometry of tetrahedra}

\begin{wrapfigure}{r}{2.75 cm}
\includegraphics[width=2.75 cm]{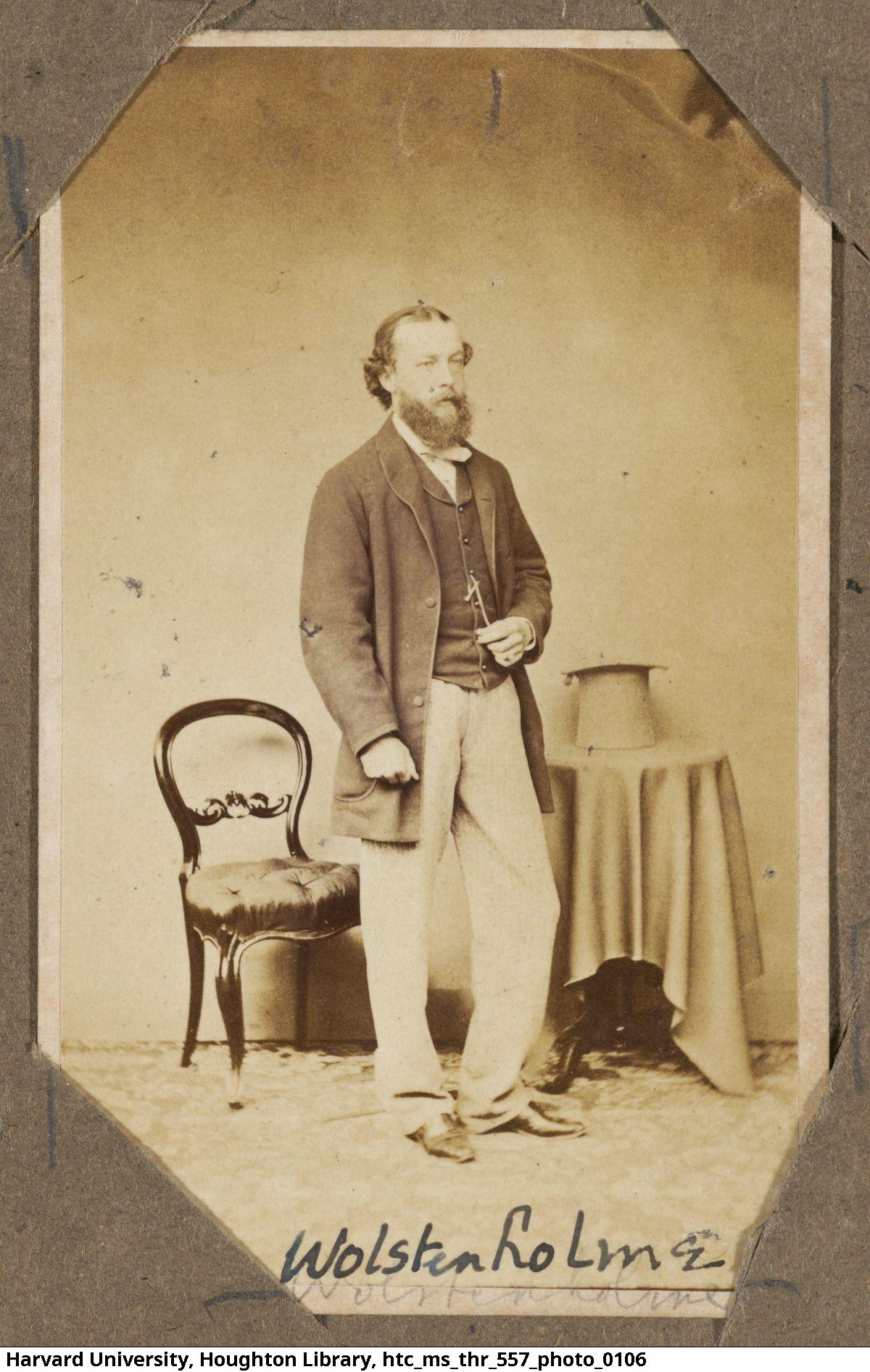}
\end{wrapfigure} 

Joseph Wolstenholme was born in 1829 in Eccles near Salford, Lancashire, England. He graduated from Cambridge in 1850 and was elected a fellow of Christ’s College in 1852. For the next twenty years he served as an examiner for the Mathematical Tripos and became rather well-known as a creator of ingenious mathematical problems. At the age of 40, Wolstenholme had to resign the fellowship because he wished to get married, and fellows at Cambridge University were historically not allowed to marry due to the institution's medieval clerical origins, where celibacy was required. After a few years of poverty, he settled as the first professor of mathematics at the newly created Royal Indian Engineering College. Shortly after turning 60, he was forced to retire because of a disagreement about the level of abstraction appropriate for a mathematics course aimed at engineers. Joseph Wolstenholme passed away in 1891, leaving a widow and four sons. 

In the last decade of his life, Wolstenholme became increasingly interested in the trigonometry of tetrahedra. He published his observations as problems in the mathematical section of the monthly journal {\it The Educational Times}, see the database \url{woollymathematics.com}.


\begin{figure}[H]
\includegraphics[scale=0.5]{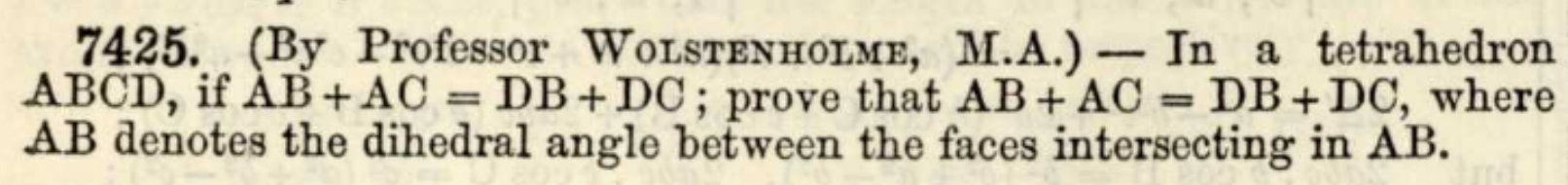}
  \caption{Problem 7425. {\it Educational Times}, Vol. 36, August 1883.}
  \label{Fig7425}
\end{figure}

Since his years in Cambridge, Wolstenholme had been one of the most active contributors to the journal. The sequence of problems relevant to us begins at Problem 7425, see Fig. \ref{Fig7425} (for some reason, the same notation is used for the lengths of edges and dihedral angles). This problem caught my attention, because it can be interpreted and solved using the connection with rational elliptic surfaces from \cite{Rud22} as both conditions $AB+AC=DB+DC$ correspond to the existence of an extra reducible fiber, resulting in the vanishing of both period maps on a root of the $D_6$-lattice. 

\begin{figure}[H]
\includegraphics[scale=0.6]{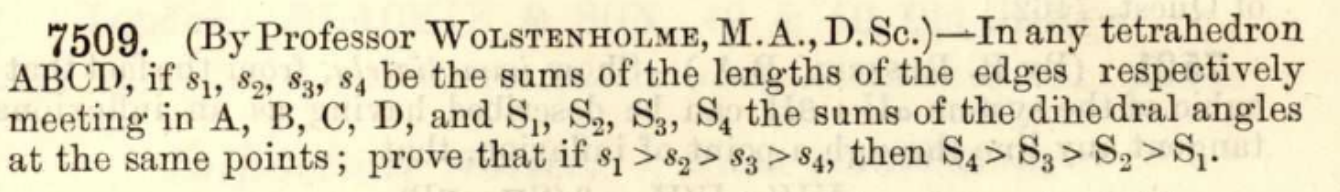}
  \caption{Problem 7509. {\it Educational Times}, Vol. 36, November 1883.}
  \label{Fig7509}
\end{figure}

A few months later,  Problem 7509 was published (Fig. \ref{Fig7509}), which can be viewed as a `soft' version of $(\ref{FormulaCrossRatios})$.

\begin{figure}[H]
\hbox{ \includegraphics[scale=0.5]{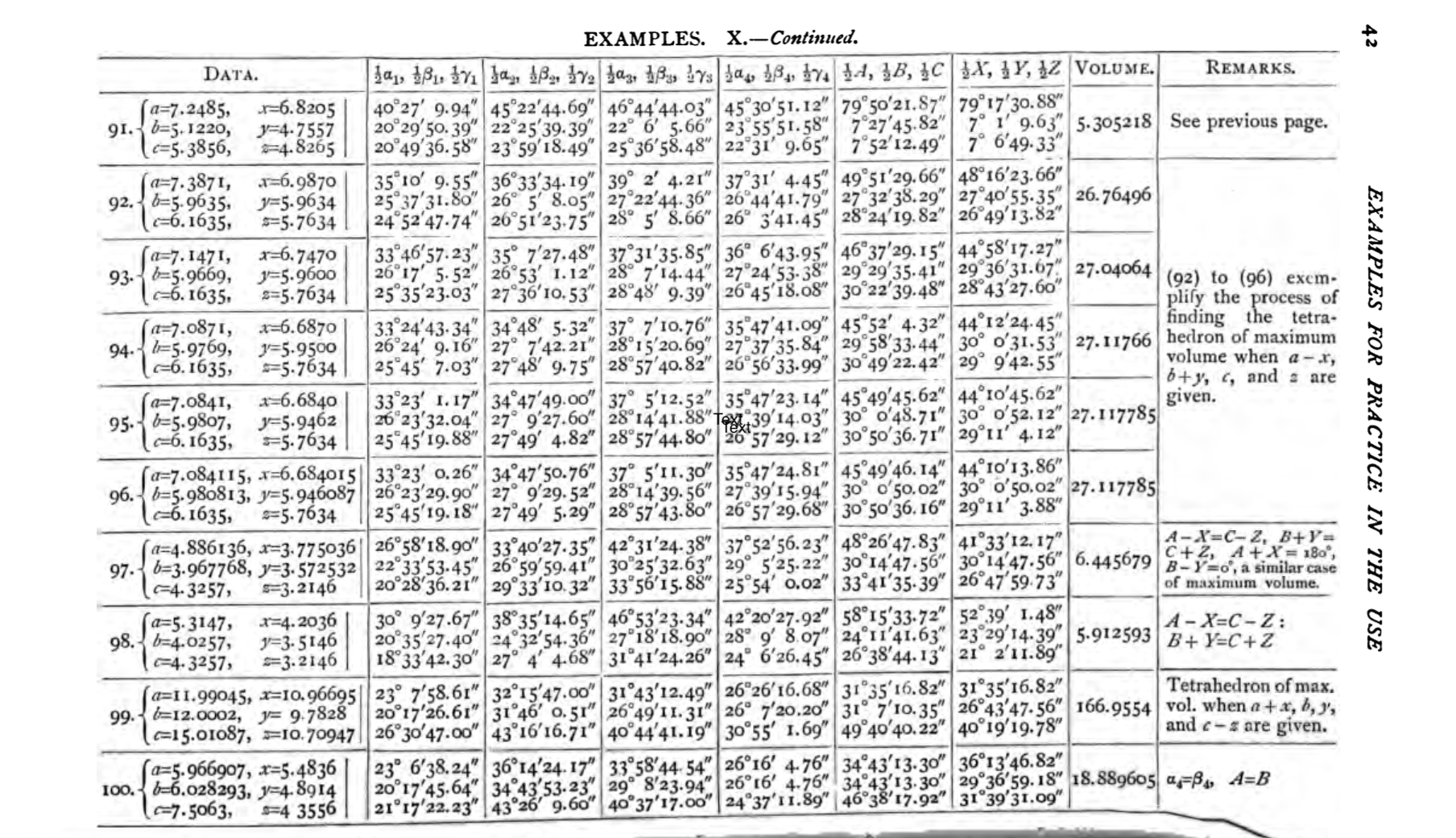}}
  \caption{Examples for Practice in the Use of Seven-figure Logarithms, p. 41.}
  \label{FigExamplesOfPractice}
\end{figure}

In the course of the next decade, 16 more problems appeared in the journal, all of which were related to the trigonometry of tetrahedra. In 1888, shortly before retiring, Wolstenholme published a book called {\it Examples for Practice in the Use of Seven-figure Logarithms} \cite{Wol88}. Traditionally, students were taught to use logarithms in computations on examples from the trigonometry triangles. Wolstenholme discusses 225 similar examples based on tetrahedra! To these examples, he adds ``remarks,'' often special cases of the theorems he had discovered. For example, on p. 40 (Fig. \ref{FigExamplesOfPractice}) Wolstenholme presents ten tetrahedra, their lengths of edges, plain and dihedral angles, and volumes.

In November 1888, Wolstenholme sought to publish a note “Certain Algebraic Results Deduced from the Geometry of the Quadrangle and Tetrahedron” but it was rejected by the referees. In the next three years, another three `tetrahedral' problems appeared in the journal. Wolstenholme passed away in November 1891.

\section{Wolstenholme's Theorem}
\begin{figure}[H]
\includegraphics[scale=0.70]{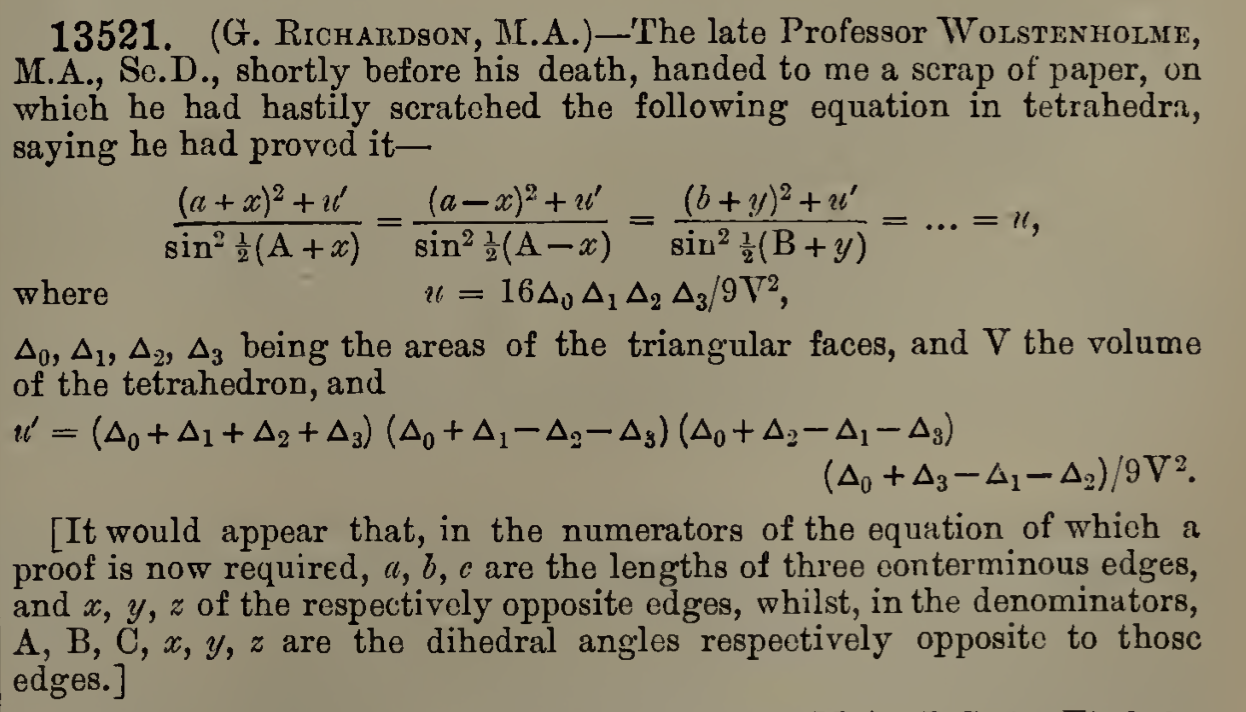}
  \caption{Problem 13521. {\it Educational Times}, Vol. 50, June 1897.}
  \label{Fig13521}
\end{figure}

In June 1897 the following problem (Fig. \ref{Fig13521}) appeared in {\it The Educational Times} authored by George Richardson, a friend of Wolstenholme. 

Here is the statement of the theorem in more standard notation. 
\begin{theorem}[J. Wolstenholme, 1889]\label{TheoremWolstenholme} Let $T$ be a Euclidean tetrahedron with lengths of edges $l_{ij}$ and dihedral angles $\alpha_{ij}.$ Then there exist number $\lambda\in \mathbb{R}$ such that
\begin{align*}
&\frac{\cos(\alpha_{12}+\alpha_{34})}{(l_{12}+l_{34})^2+\lambda}= 
\frac{\cos(\alpha_{12}-\alpha_{34})}{(l_{12}-l_{34})^2+\lambda}= 
\frac{\cos(\alpha_{13}+\alpha_{24})}{(l_{13}+l_{34})^2+\lambda} \\
=&\frac{\cos(\alpha_{13}-\alpha_{24})}{(l_{13}-l_{24})^2+\lambda}= 
\frac{\cos(\alpha_{14}+\alpha_{23})}{(l_{14}+l_{23})^2+\lambda}= 
\frac{\cos(\alpha_{14}-\alpha_{23})}{(l_{14}-l_{23})^2+\lambda}.
\end{align*}
\end{theorem}


It took me some time to notice that Theorem \ref{TheoremWolstenholme} is equivalent to (\ref{FormulaCrossRatios}). Indeed, Theorem \ref{TheoremWolstenholme}  is equivalent to the identity 
\begin{align}\label{FormulaCrossRatiosW}
[\cos(\alpha_{14}-\alpha_{23}),\cos(\alpha_{13}-\alpha_{24}),\cos(\alpha_{12}-\alpha_{34}),\infty]=[(l_{14}-l_{23})^2,(l_{13}-l_{24})^2,(l_{12}-l_{34})^2,\infty].
\end{align}
Notice that 
\[
\Bigl[\frac{a+a^{-1}}{2},\frac{b+b^{-1}}{2},\frac{c+c^{-1}}{2},\infty\Bigr]=[1,ab,ac,bc],
\]
so 
\[
[\cos(\alpha_{14}-\alpha_{23}),\cos(\alpha_{13}-\alpha_{24}),\cos(\alpha_{12}-\alpha_{34}),\infty]=\bigl[e^{i\Omega_1},e^{i\Omega_2},e^{i\Omega_3},e^{i\Omega_4}\bigr].
\]
Similarly,
\[
\bigl[a^2,b^2,c^2,\infty]=[0,a+b,a+c,b+c],
\]
so
\[
[(l_{14}-l_{23})^2,(l_{13}-l_{24})^2,(l_{12}-l_{34})^2,\infty]=[P_1,P_2,P_3,P_4].
\]
We conclude that (\ref{FormulaCrossRatiosW}) coincides with (\ref{FormulaCrossRatios}).

Theorem 1.1 from \cite{Rud22} implies other identities of cross-ratios, for instance,
\[
\bigl[e^{i\Omega_1},e^{i\Omega_2},e^{i\Omega_3},1\bigr]=[P_1,P_2,P_3,0].
\]
It is not hard to see that the full set of cross-ratio invariants obtained from \cite[Theorem 1.1]{Rud22} defines the tetrahedron up to an isometry, so \cite[Theorem 1.1]{Rud22} is stronger than Theorem \ref{TheoremWolstenholme}.

I recently obtained a copy of Wolstenholme's note for the LMS, from which it is clear that in 1888 he had already discovered a special case of Theorem \ref{TheoremWolstenholme} for a tetrahedron with the sum of areas of faces $123$ and $234$ equal to the sum of areas of faces $134$ and $124.$ For such tetrahedra the statement of the theorem simplifies since $u'=0.$ 

\begin{figure}[H]
\includegraphics[scale=0.65]{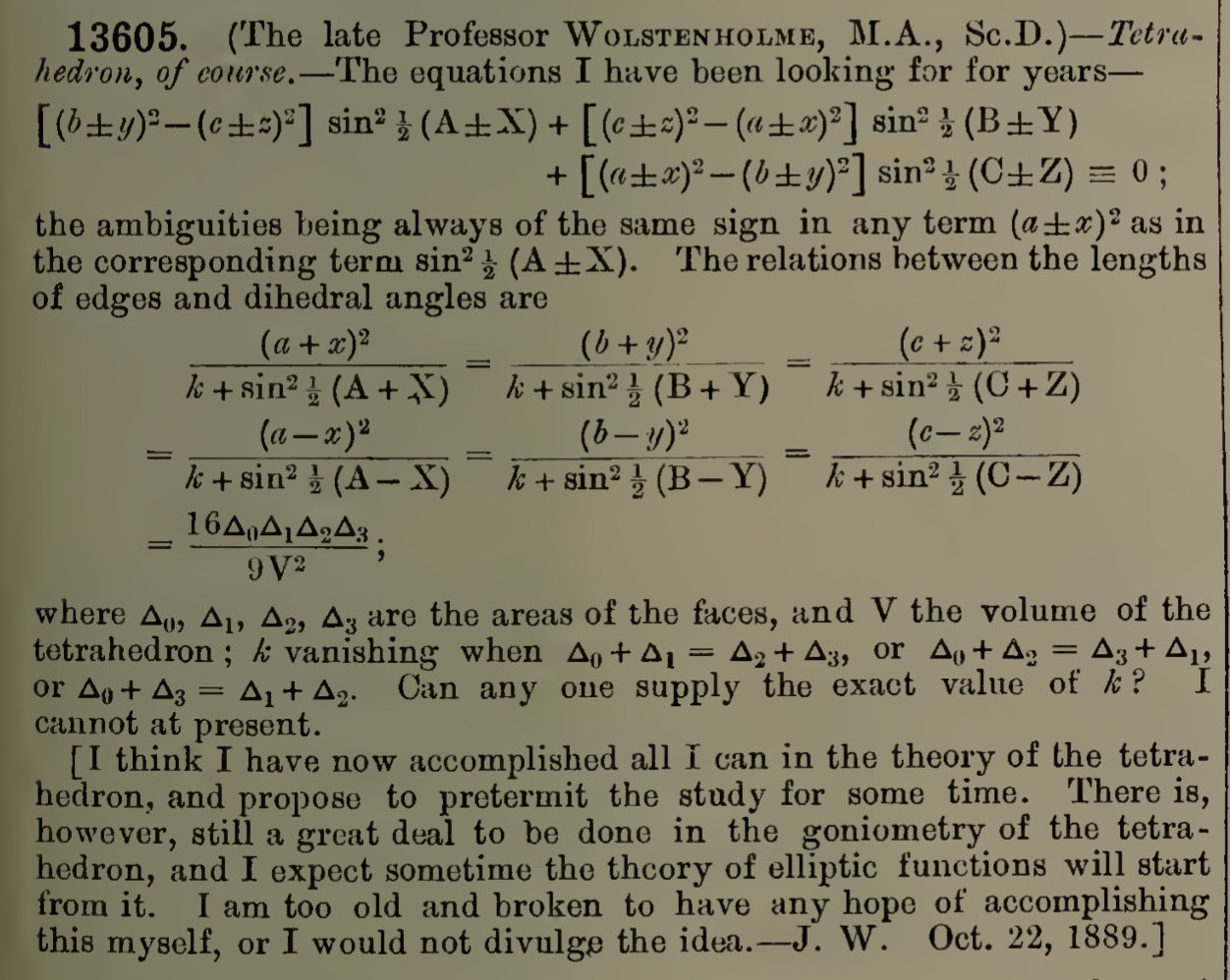}
  \caption{Problem 13605. {\it Educational Times}, Vol. 50, October 1897.}
  \label{Fig13605}
\end{figure}

A few months after Problem 13521 another problem appears in {\it The Educational Times} with the same theorem. Here, the statement is followed by a comment by Joseph Wolstenholme, see Fig. \ref{Fig13605}.

I remain curious about the specific connection to the theory of elliptic functions that Wolstenholme had in mind. My best guess is that he proposed using elliptic functions to make the statement of Theorem \ref{TheoremWolstenholme} look similar to the Law of Sines by getting rid of the constant $k.$ I cannot find any connection between that and an elliptic surface corresponding to the tetrahedron, so it seems to be ``one of those dazzling coincidences that logicians loathe and poets love.''

\section{The `Woolly One'}
We have another perspective on the years Wolstenholme spent working on tetrahedral trigonometry. Leslie Stephen, his close friend from Cambridge, used to invite him to spend summers with his family at St. Ives. In  1895 Leslie Stephen wrote in his autobiographical {\it Mausoleum Book}, \cite[p.79]{Ste77}:{\it ``I think especially of poor old Wolstenholme, called 'the woolly' by you irreverent children, a man whom I had first known as a brilliant mathematician at Cambridge, whose Bohemian tastes and heterodox opinions had made a Cambridge career inadvisable, who tried to become a hermit at Wastdale. He had emerged, married an uncongenial and rather vulgar Swiss girl, and obtained a professorship at Cooper's Hill. His four sons were badly brought up: he was despondent and dissatisfied and consoled himself with mathematics and opium. I liked him or rather was very fond of him, partly from old association and partly because feeble and faulty as he was, he was thoroughly amiable and clung to my friendship pathetically. His friends were few and his home life wretched. ... [We] had him stay every summer with us in the country. There at least he could be without his wife.''}

\begin{wrapfigure}{r}{6 cm}
\includegraphics[width=6 cm]{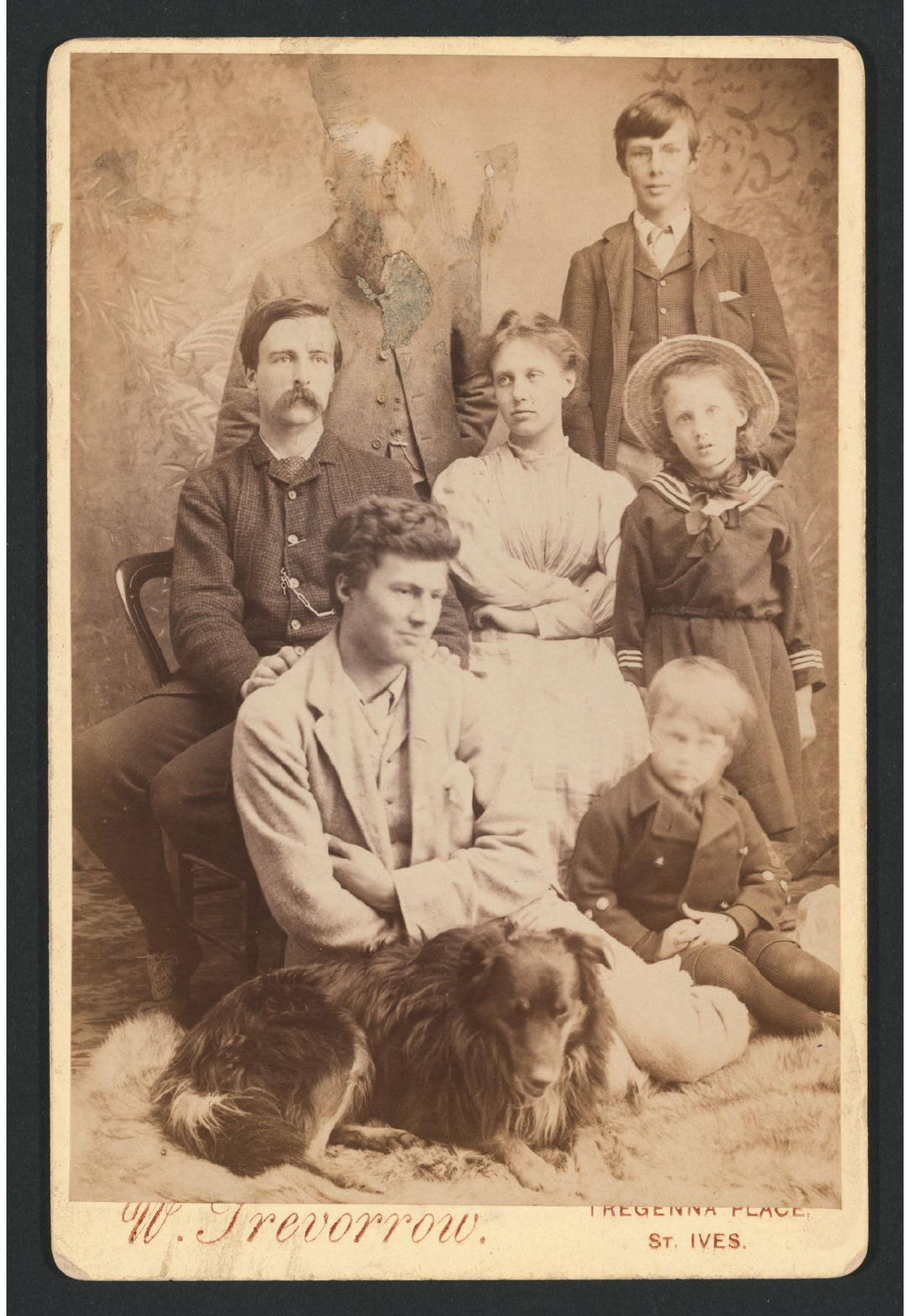}
   \caption{{\small {\it Back row left:} Joseph Wolstenholme; {\it middle row right:} Virginia Stephen (Woolf). St. Ives, ca. 1889.}}
\end{wrapfigure} 

Leslie Stephen's daughter Virginia Woolf was born in 1882, and Wolstenholme was her father's summer guest during her childhood. He made an impression on her, and she left a compelling child's eye view of him in the autobiographical essay {\it A Sketch of the Past}, \cite{Woo85}: {\it ``I name these three people because they all died when I was a child. Therefore they have never been altered. I see them exactly as I saw them then. Mr Wolstenholme was a very old gentleman who came every summer to stay with us. He was brown; he had a beard and very small eyes in fat cheeks; and he fitted into a brown wicker beehive chair as if it had been his nest. He used to sit in this beehive chair smoking and reading. He had only one characteristic
-- that when he ate plum tart he spurted the juice through his nose so that it made a purple stain on his grey moustache. This seemed enough to cause us perpetual delight. We called him ``The Woolly One''. By way of shading him a little I remember that we had to be kind to him because he was not happy at home; that he was very poor, yet once gave Thoby half a crown; that he had a son who was drowned in Australia; and I know too that he was a great mathematician. He never said a word all the time I knew him. But he still seems to me a complete character; and whenever I think of him I begin to laugh.''}

Joseph Wolstenholme became the prototype of Augustus Carmichael, \cite[I, p.32]{Bel72}, the only character in {\it To The Lighthouse} whose talent is not questioned in the novel. I find it astonishing that the powerful image of Carmichael was inspired by Wolstenholme whom the writer observed throughout the entire process of his discovery of the trigonometry of tetrahedra.

\pagebreak

\bibliographystyle{abbrv}      
\bibliography{Woolly} 

\end{document}